\documentclass[aps,prl,twocolumn,showcaps,superscriptaddress,showpacs,amsmath,amssymb]{revtex4}

\usepackage{graphicx}
\usepackage{dcolumn}

\newcommand{\kT}{k_\mathrm{B}T}
\newcommand{\vs}{\mean{\dot x}}
\newcommand{\js}{j_\mathrm{s}}
\newcommand{\ps}{p_\mathrm{s}}
\newcommand{\vloc}{v_\mathrm{s}}
\newcommand{\mean}[1]{\langle #1 \rangle}
\newcommand{\pd}[2]{\frac{\partial #1}{\partial #2}}
\newcommand{\IInt}[3]{\int_{#2}^{#3}\mathrm{d}#1\;}
\newcommand{\unit}[1]{\;\mathrm{#1}}

\begin{document}

\title{The Einstein relation generalized to non-equilibrium}

\author{V. Blickle}
\altaffiliation{These authors contributed equally to this work.}
\affiliation{2. Physikalisches Institut, Universit\"at Stuttgart,
Pfaffenwaldring 57, 70550 Stuttgart, Germany}

\author{T. Speck}
\altaffiliation{These authors contributed equally to this work.}
\affiliation{{II.} Institut f\"ur Theoretische Physik,
Universit\"at Stuttgart, Pfaffenwaldring 57, 70550 Stuttgart,
Germany}

\author{C. Lutz}
\affiliation{2. Physikalisches Institut, Universit\"at Stuttgart,
Pfaffenwaldring 57, 70550 Stuttgart, Germany}

\author{U. Seifert}
\affiliation{{II.} Institut f\"ur Theoretische Physik,
Universit\"at Stuttgart, Pfaffenwaldring 57, 70550 Stuttgart,
Germany}

\author{C. Bechinger}
\affiliation{2. Physikalisches Institut, Universit\"at Stuttgart,
Pfaffenwaldring 57, 70550 Stuttgart, Germany}

\begin{abstract}
  The Einstein relation connecting the diffusion constant and the mobility is
  violated beyond the linear response regime. For a colloidal particle driven
  along a periodic potential imposed by laser traps, we test the recent
  theoretical generalization of the Einstein relation to the non-equilibrium
  regime which involves an integral over measurable velocity correlation
  functions.
\end{abstract}


\maketitle


A comprehensive theory of systems driven out of equilibrium is still lacking
quite in contrast to the universal description of equilibrium systems by the
Gibbs-Boltzmann distribution. Linear response theory provides exact relations
valid, however, only for small deviations from equilibrium~\cite{kubo}. The
arguably most famous linear response relation is the Einstein relation
\begin{equation}
  \label{eq:fdt}
  D = \kT \mu,
\end{equation}
involving the diffusion constant $D$, the mobility $\mu$, and the thermal
energy $\kT$~\cite{eins05}. In his original derivation for a suspension in a
force field, Einstein balances the diffusive current with a linear drift. The
Einstein relation embodies a deep connection between fluctuations causing
diffusion and dissipation responsible for friction expressed by a finite
mobility.

In the present Letter, we report on the extension of the classical Einstein
relation beyond the linear response regime using a driven colloidal particle
as a paradigmatic system. Our previous theoretical work~\cite{spec06} and its
present experimental test thus introduce a third type of exact relation valid
for and relevant to small driven systems coupled to a heat bath of constant
temperature $T$. The previously discovered exact relations comprise, first,
the fluctuation theorem~\cite{evan93,gall95} which quantifies the steady state
probability of observing trajectories of negative entropy production. Second,
the Jarzynski relation~\cite{jarz97} expresses the free energy difference
between different equilibrium states by a nonlinear average of the work spent
in driving such a transition~\cite{bust05}. Both the fluctuation theorem and
the Jarzynski relation as well as their theoretical
extensions~\cite{croo00,hata01,seif05a} have been tested in various
experimental systems such as micro-mechanically manipulated
biomolecules~\cite{liph02,coll05}, colloids in time-dependent laser
traps~\cite{wang02,blic06,trep04}, Rayleigh-Benard convection~\cite{cili98},
mechanical oscillators~\cite{doua05}, and optically driven single two-level
systems~\cite{schu05}. Such exact relations (and the study of their
limitations) are fundamentally important since they provide the first elements
of a future more comprehensive theory of non-equilibrium systems.


For a non-equilibrium extension of the Einstein relation~(\ref{eq:fdt}),
consider the overdamped motion $x(t)$ of a particle moving along a periodic
one-dimensional potential $V(x)$ governed by the Langevin equation
\begin{equation}
  \dot x(t)=\frac{1}{\gamma} F(x(t))+\xi (t)
  \label{eq:langevin}
\end{equation}
with $F=-\partial V/ \partial x+f$ and $f$ a non-conservative force. The
friction coefficient $\gamma$ determines the correlations
$\mean{\xi(t)\xi(t')}=2(\kT/\gamma)\delta(t-t')$ of the white noise $\xi$.
Therefore Eq.~\eqref{eq:langevin} describes a colloidal bead driven to
non-equilibrium under the assumption that the fluctuating forces arising from
the heat bath are not affected by the driving.

For the crucial quantities $D$ and $\mu$, it is convenient to adapt
definitions which can be used both in equilibrium and beyond linear response,
i.e., in a non-equilibrium steady state characterized by $f=\text{const.}\neq
0$. The diffusion coefficient is given by
\begin{equation}
  D=\lim_{t\to \infty}\,[\mean{x^2(t)}-\mean{x(t)}^2]/(2t),
\label{eq:diff}
\end{equation}
where $\mean{\cdots}$ denotes the ensemble average. Both theoretical
work~\cite{reim01} and a recent experiment~\cite{lee06} have shown that the
force-dependent diffusion constant can be substantially larger than its
equilibrium value. The mobility
\begin{equation}
  \mu = \pd{\vs}{f}
\end{equation}
quantifies the response of the mean velocity $\vs$ to a small change of the
external force $f$. If the response is taken at $f=0$, which corresponds to
equilibrium, one has the linear response relation~(\ref{eq:fdt}).  How does
the Einstein relation change for $f\neq0$, i.e., what is the relation between
a force-dependent diffusion constant $D(f)$ and a force-dependent mobility
$\mu(f)$? Is there a simple relation at all? We have recently shown that under
non-equilibrium conditions the Einstein relation~(\ref{eq:fdt}) has to be
replaced by~\cite{spec06}
\begin{equation}
  \label{eq:fdr}
  D = \kT\mu + \IInt{\tau}{0}{\infty}I(\tau),
\end{equation}
where the second term on the right hand side is given by an integral over a
known ``violation function'' $I(\tau)$ involving measurable velocity
correlations to be discussed in detail below. Such a relation is complementary
to introducing an effective temperature which replaces $T$ in
Eq.~(\ref{eq:fdt}) in an attempt to keep its simple form~\cite{cris03,haya04}.
It has the advantage that knowledge of $I(\tau)$ offers us a better
understanding of the crucial characteristics of the non-equilibrium steady
state that causes the breakdown of the Einstein relation~\eqref{eq:fdt}.


\begin{figure}
  \includegraphics[width=0.8\linewidth]{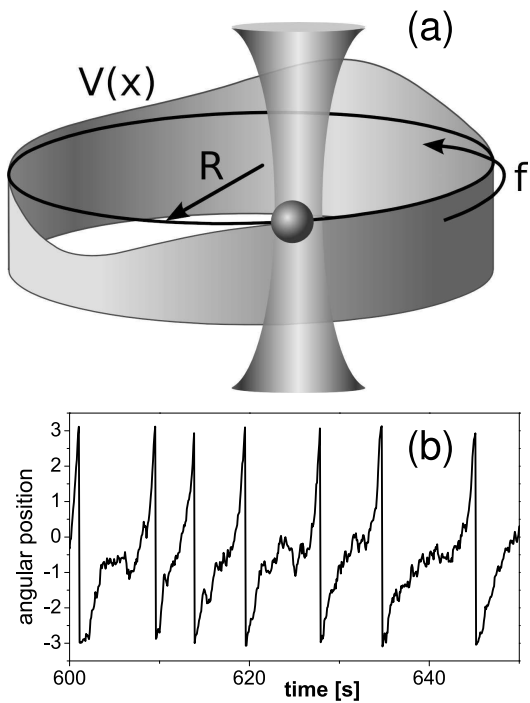}
  \caption{(a) Experimental setup. (b) Typical trajectory of the angular
    particle position for a mean particle revolution
    time~$\simeq5.8\unit{s}$.}
\end{figure}

In our experiment we subject a single colloidal silica bead with
$1.85\unit{\mu m}$ diameter to a non-equilibrium steady state by forcing it
along a toroidal trap ($R=1.75\unit{\mu m}$) created by tightly focussed
rotating optical tweezers ~\cite{fauc95,lutz06} (see Fig.~1). This is achieved
by focusing the beam of a Nd:YAG laser ($\lambda = 532\unit{nm}$) with a
microscope objective (100x, NA=1.3) into a sample cell containing a highly
diluted aqueous suspension of silica particles with $1.85\unit{\mu m}$
diameter. A pair of galvanometric driven mirrors with telescope optics
deflects the beam along a circular path and thus confines the silica bead to
an effectively one-dimensional motion. Depending on the velocity of the
rotating trap three different regimes can be distinguished~\cite{fauc95}. (i)
For small velocities friction forces are much smaller than the trapping force,
the trapped particle is able to follow the trap. (ii) With increasing velocity
the trap is not strong enough to compensate the viscous force of the fluid,
the particle escapes from the laser trap. However, every time the laser passes
the particle it is still dragged a small distance along the circle and moves
with a constant mean velocity around the torus. (iii) As the focus speed
increases (quasi)-equilibrium conditions are established and the particle is
able to diffuse freely along the torus.  With the trap rotation frequency set
to $567\unit{Hz}$ the experiments are performed in the intermediate regime
(ii) where the particle is observed to circulate with a constant mean
velocity.  Since the displacement of the particle by a single kick depends on
the laser intensity and is approximately $10 \unit{nm}$, under our
experimental conditions the spatial ($50\unit{nm}$) and temporal
($80\unit{ms}$) resolution of digital video microscopy is not sufficient to
resolve single ''kicking'' events. Therefore the particle can be considered to
be subjected to a constant force $f$ along the angular direction $x$.
Additionally the scanning motion is synchronized with an electro-optical
modulator (EOM) which allows the periodic variation of the laser intensity
along the toroid. In the experiment the tweezers intensity $P$ is weakly
modulated ($\Delta P / P \leq 10\%$). This small intensity modulation
superimposes an additional periodic potential $V(x)$ acting on the particle
when moving along the torus. As the result, the particle moves in a tilted
periodic potential. Both the potential $V(x)$ and the driving force $f$ are
not known from the input values to the EOM but must be reconstructed as
described in detail below.

\begin{figure}
  \includegraphics[width=\linewidth]{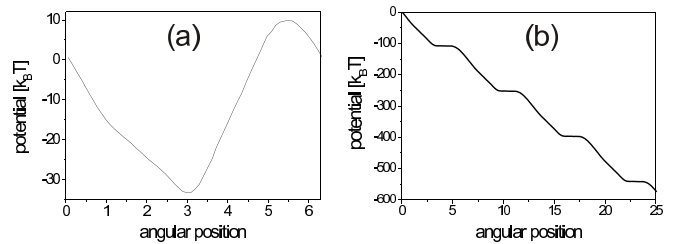}
  \caption{ (a) Reconstructed potential $V(x)$. (b) Tilted potential. The
    colloidal particle is subjected to a constant driving force
    $f\simeq0.06\unit{pN}$ and the periodic potential $V(x)$.}
\end{figure}


The central quantitity of Eq.~(\ref{eq:fdr}) is the violation function
$I(\tau)$ which can be written as~\cite{spec06}
\begin{equation}
  \label{eq:vio}
  I(\tau) = \mean{[\dot x(t+\tau)-\vs][\vloc(x(t))-\vs]}.
\end{equation}
It correlates the {\it actual} velocity $\dot x(t)$ with the {\it local} mean
velocity $\vloc(x)$ subtracting from both the {\it global} mean velocity
$\vs=2\pi R\js$ that is given by the net particle flux $\js$ through the
torus. In one dimension for a steady state, the current must be the same
everywhere and hence $\js$ is a constant. The offset $t$ is arbitrary because
of time-translational invariance in a steady state and in the following we set
$t=0$. The local mean velocity $\vloc(x)$ is the average of the stochastic
velocity $\dot x$ over the subset of trajectories passing through $x$. An
equivalent expression is $\js=\vloc(x)\ps(x)$ connecting the current with the
probability density $\ps(x)$. The local mean velocity can thus be regarded as
a measure of the local violation of detailed balance. Since in equilibrium
detailed balance holds and therefore $\vloc(x)=\vs=0$, the
violation~(\ref{eq:vio}) vanishes and Eq.~(\ref{eq:fdr}) reduces to
Eq.~(\ref{eq:fdt}).


\begin{figure*}[t]
  \includegraphics[width=0.8\linewidth]{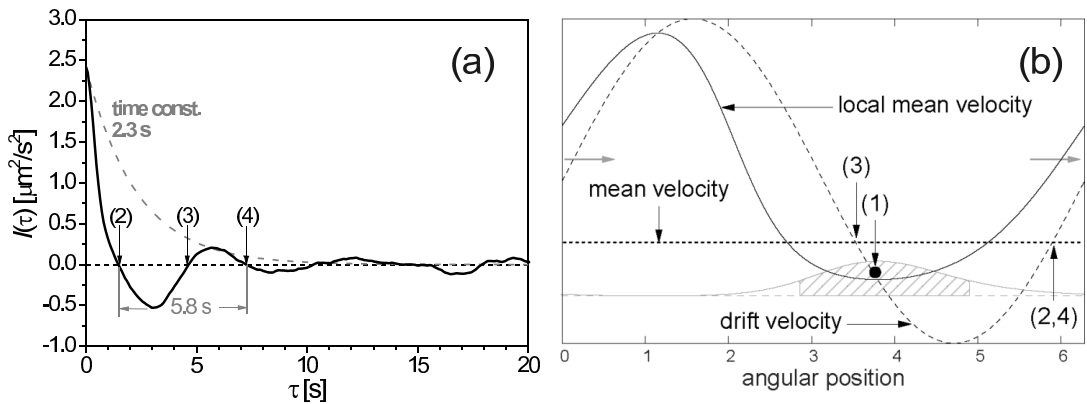}
  \caption{a) Experimentally measured violation function $I(\tau)$ (solid
    line). b) Comparison of the velocities involved in the violation function
    $I(\tau)$. For an ideal cosine potential, we sketch the probability
    distribution $\ps(x)$ (solid gray line), the local mean velocity
    $\vloc(x)$ together with the drift velocity and their mean $\vs$ versus
    the angular particle position. The drift velocity is the deterministic
    part $F/\gamma$ of the actual velocity $\dot x$. The sign change in
    $I(\tau)$ at (2), (3), and (4) can be understood as follows. In a steady
    state, a single particle trajectory will start with highest probability in
    the shaded region and, for an illustration, we choose its maximum as
    starting point~(1) determining the value $\vloc(x(t))$ in
    Eq.~(\ref{eq:vio}). Neglecting thermal fluctuations, the particle would
    follow the dashed line and during a small time step $\tau$ the product
    $F(x(t+\tau))\vloc(x(t))$ is positive. If the particle passes~(2), the
    product would become negative. The sign changes again if the particle
    passes~(3) and then~(4) and so on due to the periodic nature of the
    potential. Thermal noise and averaging over all trajectories does not
    change this behavior responsible for the oscillations of $I(\tau)$.}
\end{figure*}

For an experimental test of the non-equilibrium Einstein
relation~(\ref{eq:fdr}), we measure trajectories of a single colloidal
particle for different driving forces $f$ by adjusting the intensity
transmitted through the EOM. From a linear fit to the data we first determine
the mean global velocity $\vs$. Next, we extract the mean local velocity
$\vloc(x)$ from the histogram $\ps(x)$ with the coordinate $x$ confined to
$0\leq x\leq 2 \pi R$. Since measurements are performed with a sampling rate
of $80\unit{ms}$, we cannot directly access the velocity $\dot x(t)$
experimentally. To calculate the violation integral $I(\tau)$, we decompose
$\dot x(t)$ into a randomly fluctuating Brownian part and a drift term, see
Eq.~\eqref{eq:langevin}. We then transform $I(\tau)$ as
\begin{equation}
  \begin{split}
    I(\tau) = &\mean{[\vloc(x(\tau))-\frac{\kT}{\gamma}\pd{\Phi}{x}(x(\tau))]
      \vloc(x(0))}-\vs^2\\ &+\mean{\xi(\tau)\vloc(x(0))}.
  \end{split}
\end{equation}
The generalized potential $\Phi(x)$ is determined via the measured stationary
probability distribution, $\ps(x)=\exp[-\Phi(x)]$~\cite{spec06}. For $\tau>0$,
the last term vanishes because then $\xi(t+\tau)$ and $x(t)$ are uncorrelated.
Thus the function $I(\tau)$ depends on two measurable quantities, the current
$\js$ and the stationary probability distribution $\ps(x)$.

The potential $V$ and the driving force $f$ are determined by integrating the
force
\begin{equation}
  F = -\pd{V}{x} + f = \gamma \vloc - \kT\pd{\Phi}{x}
\end{equation}
along the torus. We obtain
\begin{equation}
  \label{eq:force}
  f = \frac{\gamma}{2\pi R}\IInt{x}{0}{2\pi R}\vloc(x)
\end{equation}
and
\begin{equation}
  V(x) = \kT\Phi(x) + \IInt{x'}{0}{x}[f- \gamma\vloc(x')]
\end{equation}
up to an irrelevant constant. In Eq.~\eqref{eq:force}, terms involving $V$ and
$\Phi$ are zero due to the periodicity of our system. Both, the potential
$V(x)$ and the tilted potential $V(x)-fx$ are shown in Fig.~2. The mobility
$\mu = \Delta\vs / \Delta f$ is determined from the change of the global mean
velocity $\Delta\vs$ upon a small variation of the force $\Delta f$.

With the experimentally determined quantities, we measure the violation
function $I(\tau)$ shown as solid line in Fig.~3a for $f=0.06\unit{pN}$. It
clearly displays the two time scales present in the system. First, the driving
leads to an oscillatory behavior with a period equal to the mean revolution
time~$\simeq5.8\unit{s}$. Second, the diffusion causes a broadening of the
particle's position resulting in a decorrelation between actual and local
velocity and hence an exponential decay with time constant $\simeq2.3\unit{s}$
indicated by the dashed line (Fig.~3a). To understand the behavior of $I(\tau)$
in more detail it is helpful to compare the different velocities involved in
the violation function $I(\tau)$ which are sketched in Fig.~3b.

After numerical integration of the experimentally determined $I(\tau)$ we
finally calculate the diffusion coefficient according to Eq.~\eqref{eq:fdr}.
To quantify the relative importance of the violation integral we plot the two
terms of the right hand side of Eq.~\eqref{eq:fdr} separately for five
different values of the driving force in Fig.~4. Their sum is in good
agreement with the independently measured diffusion coefficient directly
obtained from the particles trajectory using Eq.~\eqref{eq:diff}. As the
maximal error for the independent measurements we estimated from our data
$\pm3\%$ for the diffusion coefficient $D$, up to $\pm10\%$ for the violation
integral, and $\pm7\%$ for the mobility $\mu$.

\begin{figure}
  \includegraphics[width=\linewidth]{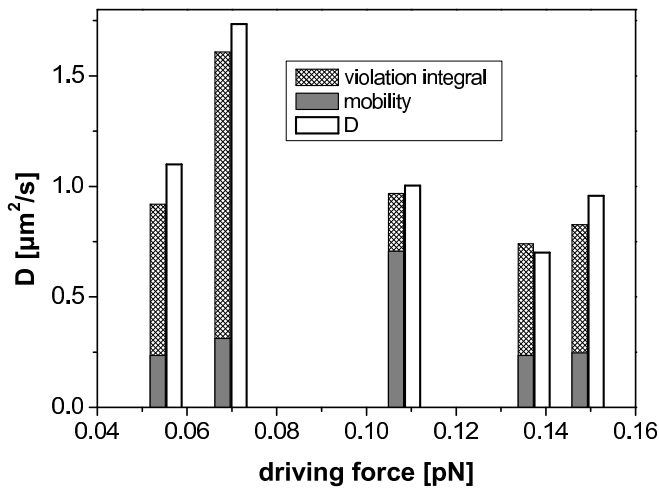}
  \caption{Experimental test of Eq.~(\ref{eq:fdr}) for different driving
    forces $f$. The open bars show the measured diffusion coefficients $D$.
    The stacked bars are mobility (gray bar) and integrated violation (hatched
    bar), respectively.}
\end{figure}

We emphasize that under our experimental parameters the violation term
dominates the diffusion coefficient (up to $80\%$) and must not be ignored. In
Fig.~4 one observes a non-monotonic dependence of the violation integral on
the driving force. This is due to the fact that the maxima of $\mu(f)$ and
$D(f)$ do not occur at the same driving force but are slightly
offset~\cite{reim01}. This implies for the violation function a maximum
followed by a minimum as a function of $f$. For very small driving forces, the
bead is close to equilibrium and its motion can be described using linear
response theory. As a result, the violation integral is negligible.
Experimentally, this regime is difficult to access since $D$ and $\mu$ become
exponentially small and cannot be measured at reasonable time scales for small
forces and potentials as deep as $40\,\kT$ (cf. Fig.~2a). For much larger
forces, the relative magnitude of the violation term becomes smaller as well.
In this limit, the imposed potential becomes irrelevant and the spatial
dependence of the local mean velocity, which is the source of the violation
term, vanishes.  The fact that in our regime the violation term is of the same
order of magnitude as the mobility proves that we are indeed probing the
regime beyond linear response.  Still, the description of the colloidal motion
by a Markovian (memory-less) Brownian motion with drift as implicit in our
analysis remains obviously a faithful representation since the theoretical
results are derived from such a framework.

The Einstein relation generalized to non-equilibrium as presented and tested
here for the driven motion along a single coordinate could be considered as a
paradigm. Extending such an approach to interacting particles and resolving
frequency dependent versions of Eq.~(\ref{eq:vio})~\cite{spec06} while
certainly experimentally challenging will provide further insight into crucial
elements of a future systematic theory of non-equilibrium systems.



\begin{thebibliography}{24}
\expandafter\ifx\csname
natexlab\endcsname\relax\def\natexlab#1{#1}\fi
\expandafter\ifx\csname bibnamefont\endcsname\relax
  \def\bibnamefont#1{#1}\fi
\expandafter\ifx\csname bibfnamefont\endcsname\relax
  \def\bibfnamefont#1{#1}\fi
\expandafter\ifx\csname citenamefont\endcsname\relax
  \def\citenamefont#1{#1}\fi
\expandafter\ifx\csname url\endcsname\relax
  \def\url#1{\texttt{#1}}\fi
\expandafter\ifx\csname
urlprefix\endcsname\relax\def\urlprefix{URL }\fi
\providecommand{\bibinfo}[2]{#2}
\providecommand{\eprint}[2][]{\url{#2}}

\bibitem[{\citenamefont{Kubo et~al.}(1991)\citenamefont{Kubo, Toda, and
  Hashitsume}}]{kubo}
\bibinfo{author}{\bibfnamefont{R.}~\bibnamefont{Kubo}},
  \bibinfo{author}{\bibfnamefont{M.}~\bibnamefont{Toda}}, \bibnamefont{and}
  \bibinfo{author}{\bibfnamefont{N.}~\bibnamefont{Hashitsume}},
  \emph{\bibinfo{title}{Statistical Physics II}}
  (\bibinfo{publisher}{Springer-Verlag}, \bibinfo{address}{Berlin},
  \bibinfo{year}{1991}), \bibinfo{edition}{2nd} ed.

\bibitem[{\citenamefont{Einstein}(1905)}]{eins05}
\bibinfo{author}{\bibfnamefont{A.}~\bibnamefont{Einstein}},
  \bibinfo{journal}{Ann.\ Phys.} \textbf{\bibinfo{volume}{17}},
  \bibinfo{pages}{549} (\bibinfo{year}{1905}).

\bibitem[{\citenamefont{Speck and Seifert}(2006)}]{spec06}
\bibinfo{author}{\bibfnamefont{T.}~\bibnamefont{Speck}} \bibnamefont{and}
  \bibinfo{author}{\bibfnamefont{U.}~\bibnamefont{Seifert}},
  \bibinfo{journal}{Europhys.\ Lett.} \textbf{\bibinfo{volume}{74}},
  \bibinfo{pages}{391} (\bibinfo{year}{2006}).

\bibitem[{\citenamefont{Evans et~al.}(1993)\citenamefont{Evans, Cohen, and
  Morriss}}]{evan93}
\bibinfo{author}{\bibfnamefont{D.~J.} \bibnamefont{Evans}},
  \bibinfo{author}{\bibfnamefont{E.~G.~D.} \bibnamefont{Cohen}},
  \bibnamefont{and} \bibinfo{author}{\bibfnamefont{G.~P.}
  \bibnamefont{Morriss}}, \bibinfo{journal}{Phys.\ Rev.\ Lett.}
  \textbf{\bibinfo{volume}{71}}, \bibinfo{pages}{2401} (\bibinfo{year}{1993}).

\bibitem[{\citenamefont{Gallavotti and Cohen}(1995)}]{gall95}
\bibinfo{author}{\bibfnamefont{G.}~\bibnamefont{Gallavotti}} \bibnamefont{and}
  \bibinfo{author}{\bibfnamefont{E.~G.~D.} \bibnamefont{Cohen}},
  \bibinfo{journal}{Phys.\ Rev.\ Lett.} \textbf{\bibinfo{volume}{74}},
  \bibinfo{pages}{2694} (\bibinfo{year}{1995}).

\bibitem[{\citenamefont{Jarzynski}(1997)}]{jarz97}
\bibinfo{author}{\bibfnamefont{C.}~\bibnamefont{Jarzynski}},
  \bibinfo{journal}{Phys.\ Rev.\ Lett.} \textbf{\bibinfo{volume}{78}},
  \bibinfo{pages}{2690} (\bibinfo{year}{1997}).

\bibitem[{\citenamefont{Bustamante et~al.}(2005)\citenamefont{Bustamante,
  Liphardt, and Ritort}}]{bust05}
\bibinfo{author}{\bibfnamefont{C.}~\bibnamefont{Bustamante}},
  \bibinfo{author}{\bibfnamefont{J.}~\bibnamefont{Liphardt}}, \bibnamefont{and}
  \bibinfo{author}{\bibfnamefont{F.}~\bibnamefont{Ritort}},
  \bibinfo{journal}{Physics Today} \textbf{\bibinfo{volume}{58}},
  \bibinfo{pages}{43} (\bibinfo{year}{2005}).

\bibitem[{\citenamefont{Crooks}(2000)}]{croo00}
\bibinfo{author}{\bibfnamefont{G.~E.} \bibnamefont{Crooks}},
  \bibinfo{journal}{Phys.\ Rev.\ E} \textbf{\bibinfo{volume}{61}},
  \bibinfo{pages}{2361} (\bibinfo{year}{2000}).

\bibitem[{\citenamefont{Hatano and Sasa}(2001)}]{hata01}
\bibinfo{author}{\bibfnamefont{T.}~\bibnamefont{Hatano}} \bibnamefont{and}
  \bibinfo{author}{\bibfnamefont{S.}~\bibnamefont{Sasa}},
  \bibinfo{journal}{Phys.\ Rev.\ Lett.} \textbf{\bibinfo{volume}{86}},
  \bibinfo{pages}{3463} (\bibinfo{year}{2001}).

\bibitem[{\citenamefont{Seifert}(2005)}]{seif05a}
\bibinfo{author}{\bibfnamefont{U.}~\bibnamefont{Seifert}},
  \bibinfo{journal}{Phys.\ Rev.\ Lett.} \textbf{\bibinfo{volume}{95}},
  \bibinfo{pages}{040602} (\bibinfo{year}{2005}).

\bibitem[{\citenamefont{Liphardt et~al.}(2002)\citenamefont{Liphardt, Dumont,
  Smith, Tinoco~Jr, and Bustamante}}]{liph02}
\bibinfo{author}{\bibfnamefont{J.}~\bibnamefont{Liphardt}},
  \bibinfo{author}{\bibfnamefont{S.}~\bibnamefont{Dumont}},
  \bibinfo{author}{\bibfnamefont{S.~B.} \bibnamefont{Smith}},
  \bibinfo{author}{\bibfnamefont{I.}~\bibnamefont{Tinoco~Jr}},
  \bibnamefont{and}
  \bibinfo{author}{\bibfnamefont{C.}~\bibnamefont{Bustamante}},
  \bibinfo{journal}{Science} \textbf{\bibinfo{volume}{296}},
  \bibinfo{pages}{1832} (\bibinfo{year}{2002}).

\bibitem[{\citenamefont{Collin et~al.}(2005)\citenamefont{Collin, Ritort,
  Jarzynski, Smith, Tinoco, and Bustamante}}]{coll05}
\bibinfo{author}{\bibfnamefont{D.}~\bibnamefont{Collin}},
  \bibinfo{author}{\bibfnamefont{F.}~\bibnamefont{Ritort}},
  \bibinfo{author}{\bibfnamefont{C.}~\bibnamefont{Jarzynski}},
  \bibinfo{author}{\bibfnamefont{S.}~\bibnamefont{Smith}},
  \bibinfo{author}{\bibfnamefont{I.}~\bibnamefont{Tinoco}}, \bibnamefont{and}
  \bibinfo{author}{\bibfnamefont{C.}~\bibnamefont{Bustamante}},
  \bibinfo{journal}{Nature} \textbf{\bibinfo{volume}{437}},
  \bibinfo{pages}{231} (\bibinfo{year}{2005}).

\bibitem[{\citenamefont{Wang et~al.}(2002)\citenamefont{Wang, Sevick, Mittag,
  Searles, and Evans}}]{wang02}
\bibinfo{author}{\bibfnamefont{G.~M.} \bibnamefont{Wang}},
  \bibinfo{author}{\bibfnamefont{E.~M.} \bibnamefont{Sevick}},
  \bibinfo{author}{\bibfnamefont{E.}~\bibnamefont{Mittag}},
  \bibinfo{author}{\bibfnamefont{D.~J.} \bibnamefont{Searles}},
  \bibnamefont{and} \bibinfo{author}{\bibfnamefont{D.~J.} \bibnamefont{Evans}},
  \bibinfo{journal}{Phys.\ Rev.\ Lett.} \textbf{\bibinfo{volume}{89}},
  \bibinfo{pages}{050601} (\bibinfo{year}{2002}).

\bibitem[{\citenamefont{Blickle et~al.}(2006)\citenamefont{Blickle, Speck,
  Helden, Seifert, and Bechinger}}]{blic06}
\bibinfo{author}{\bibfnamefont{V.}~\bibnamefont{Blickle}},
  \bibinfo{author}{\bibfnamefont{T.}~\bibnamefont{Speck}},
  \bibinfo{author}{\bibfnamefont{L.}~\bibnamefont{Helden}},
  \bibinfo{author}{\bibfnamefont{U.}~\bibnamefont{Seifert}}, \bibnamefont{and}
  \bibinfo{author}{\bibfnamefont{C.}~\bibnamefont{Bechinger}},
  \bibinfo{journal}{Phys.\ Rev.\ Lett.} \textbf{\bibinfo{volume}{96}},
  \bibinfo{pages}{070603} (\bibinfo{year}{2006}).

\bibitem[{\citenamefont{Trepagnier et~al.}(2004)\citenamefont{Trepagnier,
  Jarzynski, Ritort, Crooks, Bustamante, and Liphardt}}]{trep04}
\bibinfo{author}{\bibfnamefont{E.~H.} \bibnamefont{Trepagnier}},
  \bibinfo{author}{\bibfnamefont{C.}~\bibnamefont{Jarzynski}},
  \bibinfo{author}{\bibfnamefont{F.}~\bibnamefont{Ritort}},
  \bibinfo{author}{\bibfnamefont{G.~E.} \bibnamefont{Crooks}},
  \bibinfo{author}{\bibfnamefont{C.~J.} \bibnamefont{Bustamante}},
  \bibnamefont{and} \bibinfo{author}{\bibfnamefont{J.}~\bibnamefont{Liphardt}},
  \bibinfo{journal}{Proc.\ Natl.\ Acad.\ Sci.\ U.S.A.}
  \textbf{\bibinfo{volume}{101}}, \bibinfo{pages}{15038}
  (\bibinfo{year}{2004}).

\bibitem[{\citenamefont{Ciliberto and Laroche}(1998)}]{cili98}
\bibinfo{author}{\bibfnamefont{S.}~\bibnamefont{Ciliberto}} \bibnamefont{and}
  \bibinfo{author}{\bibfnamefont{C.}~\bibnamefont{Laroche}},
  \bibinfo{journal}{J.\ Phys.\ IV\ France} \textbf{\bibinfo{volume}{8 (P6)}},
  \bibinfo{pages}{215} (\bibinfo{year}{1998}).

\bibitem[{\citenamefont{Douarche et~al.}(2005)\citenamefont{Douarche,
  Ciliberto, Petrosyan, and Rabbiosi}}]{doua05}
\bibinfo{author}{\bibfnamefont{F.}~\bibnamefont{Douarche}},
  \bibinfo{author}{\bibfnamefont{S.}~\bibnamefont{Ciliberto}},
  \bibinfo{author}{\bibfnamefont{A.}~\bibnamefont{Petrosyan}},
  \bibnamefont{and} \bibinfo{author}{\bibfnamefont{I.}~\bibnamefont{Rabbiosi}},
  \bibinfo{journal}{Europhys.\ Lett.} \textbf{\bibinfo{volume}{70}},
  \bibinfo{pages}{593} (\bibinfo{year}{2005}).

\bibitem[{\citenamefont{Schuler et~al.}(2005)\citenamefont{Schuler, Speck,
  Tietz, Wrachtrup, and Seifert}}]{schu05}
\bibinfo{author}{\bibfnamefont{S.}~\bibnamefont{Schuler}},
  \bibinfo{author}{\bibfnamefont{T.}~\bibnamefont{Speck}},
  \bibinfo{author}{\bibfnamefont{C.}~\bibnamefont{Tietz}},
  \bibinfo{author}{\bibfnamefont{J.}~\bibnamefont{Wrachtrup}},
  \bibnamefont{and} \bibinfo{author}{\bibfnamefont{U.}~\bibnamefont{Seifert}},
  \bibinfo{journal}{Phys.\ Rev.\ Lett.} \textbf{\bibinfo{volume}{94}},
  \bibinfo{pages}{180602} (\bibinfo{year}{2005}).

\bibitem[{\citenamefont{Reimann et~al.}(2001)\citenamefont{Reimann, van~den
  Broeck, Linke, H\"{a}nggi, Rubi, and P\'{e}rez-Madrid}}]{reim01}
\bibinfo{author}{\bibfnamefont{P.}~\bibnamefont{Reimann}},
  \bibinfo{author}{\bibfnamefont{C.}~\bibnamefont{van~den Broeck}},
  \bibinfo{author}{\bibfnamefont{H.}~\bibnamefont{Linke}},
  \bibinfo{author}{\bibfnamefont{P.}~\bibnamefont{H\"{a}nggi}},
  \bibinfo{author}{\bibfnamefont{M.}~\bibnamefont{Rubi}}, \bibnamefont{and}
  \bibinfo{author}{\bibfnamefont{A.}~\bibnamefont{P\'{e}rez-Madrid}},
  \bibinfo{journal}{Phys.\ Rev.\ Lett.} \textbf{\bibinfo{volume}{87}},
  \bibinfo{pages}{010602} (\bibinfo{year}{2001}).

\bibitem[{\citenamefont{Lee and Grier}(2006)}]{lee06}
\bibinfo{author}{\bibfnamefont{S.-H.} \bibnamefont{Lee}} \bibnamefont{and}
  \bibinfo{author}{\bibfnamefont{D.~G.} \bibnamefont{Grier}},
  \bibinfo{journal}{Phys.\ Rev.\ Lett.} \textbf{\bibinfo{volume}{96}},
  \bibinfo{pages}{190601} (\bibinfo{year}{2006}).

\bibitem[{\citenamefont{Crisanti and Ritort}(2003)}]{cris03}
\bibinfo{author}{\bibfnamefont{A.}~\bibnamefont{Crisanti}} \bibnamefont{and}
  \bibinfo{author}{\bibfnamefont{F.}~\bibnamefont{Ritort}},
  \bibinfo{journal}{J.\ Phys.\ A:\ Math.\ Gen.} \textbf{\bibinfo{volume}{36}},
  \bibinfo{pages}{R181} (\bibinfo{year}{2003}).

\bibitem[{\citenamefont{Hayashi and Sasa}(2004)}]{haya04}
\bibinfo{author}{\bibfnamefont{K.}~\bibnamefont{Hayashi}} \bibnamefont{and}
  \bibinfo{author}{\bibfnamefont{S.}~\bibnamefont{Sasa}},
  \bibinfo{journal}{Phys.\ Rev.\ E} \textbf{\bibinfo{volume}{69}},
  \bibinfo{pages}{066119} (\bibinfo{year}{2004}).

\bibitem[{\citenamefont{Faucheux et~al.}(1995)\citenamefont{Faucheux,
  Stolovitzky, and Libchaber}}]{fauc95}
\bibinfo{author}{\bibfnamefont{L.}~\bibnamefont{Faucheux}},
  \bibinfo{author}{\bibfnamefont{G.}~\bibnamefont{Stolovitzky}},
  \bibnamefont{and}
  \bibinfo{author}{\bibfnamefont{A.}~\bibnamefont{Libchaber}},
  \bibinfo{journal}{Phys.\ Rev.\ E} \textbf{\bibinfo{volume}{51}},
  \bibinfo{pages}{5239} (\bibinfo{year}{1995}).

\bibitem[{\citenamefont{Lutz et~al.}(2006)\citenamefont{Lutz, Reichert, Stark,
  and Bechinger}}]{lutz06}
\bibinfo{author}{\bibfnamefont{C.}~\bibnamefont{Lutz}},
  \bibinfo{author}{\bibfnamefont{M.}~\bibnamefont{Reichert}},
  \bibinfo{author}{\bibfnamefont{H.}~\bibnamefont{Stark}}, \bibnamefont{and}
  \bibinfo{author}{\bibfnamefont{C.}~\bibnamefont{Bechinger}},
  \bibinfo{journal}{Europhys.\ Lett.} \textbf{\bibinfo{volume}{74}},
  \bibinfo{pages}{719} (\bibinfo{year}{2006}).

\end{thebibliography}
\end{document}